\documentclass[aps,twocolumn,prl,showpacs,letterpaper,superscriptaddress,floatfix]{revtex4}

\usepackage{graphicx,amsmath,amsfonts,amssymb}

\def\be{\begin{equation}}
\def\ee{\end{equation}}

\def\spf{(TMTSF)$_2$PF$_6$}
\def\sclo{(TMTSF)$_2$ClO$_4$}

\def\mba{\mathbf a}
\def\mbbp{\mathbf b'}

\def\mbcs{\mathbf c^*}
\def\bea{\begin{eqnarray} }
\def\eea{\end{eqnarray} }
\def\beaS{\begin{eqnarray*} }
\def\eeaS{\end{eqnarray*} }
\def\iTone{$T_1^{-1}$}
\def\se{$^{77}$Se}
\def\C{$^{13}$C}

\def\hc2{$H_{c2}$}
\def\iTone{$T_1^{-1}$}

\begin{document}
\title{On the superconducting state of the organic conductor \sclo}

\voffset=1cm

\begin{abstract}
\sclo\ is a quasi-one dimensional organic conductor and
superconductor with $T_c=1.4$K, and one of at least two Bechgaard
salts observed to have upper critical fields far exceeding the
paramagnetic limit. Nevertheless, the \se\ NMR Knight shift at low
fields reveals a decrease in spin susceptibility $\chi_s$ consistent
with singlet spin pairing. The field dependence of the spin-lattice
relaxation rate at 100mK exhibits a sharp crossover (or phase
transition) at a field $H_s\sim15$kOe, to a regime where $\chi_s$ is
close to the normal state value, even though \hc2 $\gg H_s$.

\end{abstract}


\author{J.~Shinagawa}
\affiliation{Department of Physics $\&$ Astronomy, UCLA, Los
Angeles, California 90095 USA}
\author{Y.~Kurosaki}
\affiliation{Department of Physics $\&$ Astronomy, UCLA, Los
Angeles, California 90095 USA} \affiliation{Department of Applied
Physics, University of Tokyo, Bunkyo-ku, Tokyo, 113-8656, Japan}
\author{F.~Zhang}
\affiliation{Department of Physics $\&$ Astronomy, UCLA, Los
Angeles, California 90095 USA}
\author{C.~Parker}
\affiliation{Department of Physics $\&$ Astronomy, UCLA, Los
Angeles, California 90095 USA} \affiliation{Department of Physics,
Princeton University, Princeton, NJ 08544 USA}
\author{S.~E.~Brown}
\affiliation{Department of Physics $\&$ Astronomy, UCLA, Los
Angeles, California 90095 USA}
\author{D.~J\'{e}rome}
\affiliation{Laboratoire de Physique des Solides, Universit\'{e}
de Paris, Sud (CNRS UMR 8502) Orsay, 91405, France}
\author{J.~B.~Christensen}
\author{K.~Bechgaard}
\affiliation{Department of Chemistry, H. C. \O rsted Institute,
Universitetsparken 5, 2100, Copenhagen, Denmark}

\date{\today}

\maketitle

Superconductivity in the Bechgaard salts (TMTSF)$_2$X is distinctive
for a number of reasons \cite{Ishiguro:1998,Jerome:2004}, but
particularly for the very large upper critical fields \hc2\ relative
to $T_c$ \cite{Lee:1997,Oh:2004}. When orbital suppression by
magnetic fields is avoided, then singlet-paired superconductivity is
still unstable beyond a paramagnetic pair-breaking field $H_p$
\cite{Clogston:1962} because of the difference in spin
susceptibility $\chi_s$ between the normal and superconducting
states. For $s-$wave superconductors in the weak-coupling limit,
$H_p=$(18kOe/K)$T_c$. \hc2\ has been reported greater than $90$kOe
\cite{Lee:2002} for \spf, an enhancement of more than four times
over $H_p=$22kOe for a $T_c=1.4$K. And for the isomorphic salt
\sclo, superconductivity beyond $50$kOe was recently reported
\cite{Oh:2004}.

Layered superconductors can exhibit upper critical fields
approaching $H_p$ when the magnetic field lies in the plane of the
layers. Commonly known examples include the high-$T_c$ cuprates
\cite{Vedeneev:2006}, and organic superconductors. Quasi-one
dimensional superconductors, as well as quasi-2D superconductors,
offer an opportunity for decoupling the layers from field-induced
confinement \cite{Lebed:1986b,Dupuis:1993,Lebed:1998}. Evidently,
{\it both} orbital suppression and spin pair-breaking of
superconductivity is weak for the Bechgaard salts.

Spin triplet pairing avoids paramagnetic limiting effects and is a
possible explanation for the large \hc2
\cite{Lee:1997,Dupuis:1993,Lebed:1999}. Indeed, previous NMR Knight
shift measurements in \spf\ under pressure were interpreted as
consistent with an equal-spin pairing triplet order parameter
\cite{Lee:2001,Lee:2003}. However, there are other circumstances
under which the Pauli limit is exceeded. For example, the spatially
inhomogeneous state described by Fulde and Ferrell, and
independently by Larkin and Ovchinnikov (FFLO)
\cite{FF:1964,LO:1965} has $\chi_s\ne0$, so it is not
paramagnetically limited at $H_p$. Also, superconductors with strong
spin-orbit scattering can exhibit large critical fields because
there is a net spin magnetization in a magnetic field
\cite{Ferrell:1959,Anderson:1959}. For the Bechgaard salts, the
proposal for triplet pairing is compelling for a number of reasons,
including: a phase transition to an inhomogeneous FFLO state has not
been identified, and, there is no evidence for significant
spin-orbit scattering in either \spf\ or in \sclo \cite{Lee:1997}.
And finally, the triplet pairing state is a known instability of
one-dimensional \cite{Giamarchi:1989} and quasi-1D electronic models
\cite{Podolsky:2004,Nickel:2005}.

The observation of large upper critical fields in \sclo\ motivated
us to investigate further with NMR techniques. In addition, we
wanted to push our own measurements to lower fields and lower
temperatures \cite{Shinagawa:2006}. Here, we report on \se\ NMR
Knight shift experiments in \sclo, performed at smaller magnetic
fields than the earlier work. The magnetic field was applied
precisely in the crystallographic layers close to the $\mba$- and
$\mbbp$-axes, at a strength just less than $H_0$=O(10kOe); this is
well below the observed critical fields from transport experiments
\cite{Oh:2004}. In both cases, a shift consistent with a decrease
of $\chi_s$ in the superconducting state is observed. The
experiments are interpreted to give evidence for spin-singlet
pairing at low field. The existence of lines of nodes is indicated
by the weak temperature dependence of the spin lattice relaxation
rate \cite{Takigawa:1987,Lee:2003}, though this aspect remains
controversial \cite{Belin:1997}. However, the nature of the
superconducting state at high fields remains a puzzle: as part of
this study we made measurements of \se\ longitudinal relaxation
(\iTone) over a range of magnetic fields at $T$=100mK. We observed
a significant and very sharply-defined increase in the dynamical
spin susceptibility within the superconducting state; the increase
divides a low-field regime (LSC) from a high-field one (HSC). We
discuss constraints imposed on the interpretation of the HSC by
existing data.

Two single crystals of \sclo, with the approximate dimensions
6$\times$2$\times$0.4mm$^{3}$ (denoted hereafter A) and
4$\times$2$\times$1mm$^3$ (B), were placed into NMR coils. Tuning
and matching elements of the NMR tank circuit were made outside
the cryostat ("top-tuning") so that a range of fields and
frequencies could be accessed. Sample B, grown at the \O rsted
Institute, Denmark, is 10\% \C-spin labelled on the bridge of the
TMTSF dimer; it was configured for measurements with the magnetic
field direction near to $\mbbp$. The other sample (A), grown at
UCLA, was configured for magnetic field alignment near to $\mba$.
The samples were mounted on the platform of a piezoelectric
rotator with 0.5 millidegree increments, and the rotation angle
was calibrated using two mutually orthogonal Hall sensors mounted
to the platform. Electrical contacts were silver-painted onto the
sample surfaces normal to the $\mbcs$ direction. The samples were
slow-cooled at the rate of 7mK/min through the anion ordering
transition ($T_{AO}$=24K) so as to reach the relaxed state and
onset of superconductivity at $T_c=1.4$K. The superconducting
transition was observed in the resistivity measurement and in
reflected rf power measurements, and alignment of the magnetic
field direction to lie precisely within the layers was
accomplished using the piezo rotator while probing the angular
dependence of the reflected power. NMR spectroscopic and
relaxation measurements were performed on both \se\ and \C\
nuclei.

\begin{figure}
\includegraphics[width=3in]{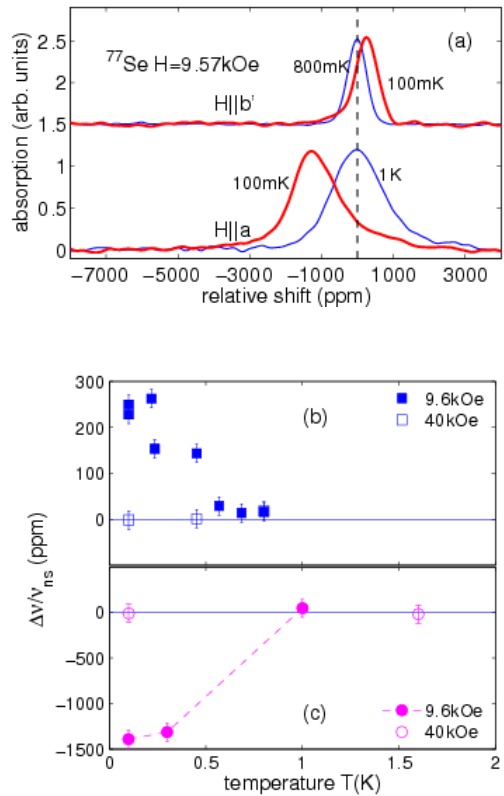}
\caption{a) \se\ spectra in the normal and superconducting states of
\sclo, for two orthogonal field orientations within the
crystallographic layers.  b), c) relative shifts vs. temperature for
$\mathbf{H}\parallel\mbbp$, $\mba$, respectively. At $H=40$kOe,
there is no observable change from the normal state shifts at the
lowest temperatures measured ($100$mK). Zero shift is arbitrarily
set to the normal state first moment. } \label{fig:seShift}
\end{figure}

In presenting the results, we start with the key observation: the
\se\ shifts in the superconducting state and normal state, and
follow with complementary characterization data: spin lattice
relaxation rates and magnetoresistance. In Fig.
\ref{fig:seShift}a, we show \se\ spectra for the two samples,
identified by the direction of the applied field. The
spectroscopic experiments were performed at small tip-angles
($<3^\cdot$) to avoid heating. More specifically, temperature
rises were detected by time-synchronous resistivity measurements,
and for both samples we were able to use sufficiently small tip
angles and associated pulse energies so that temperature rises
were undetectable. The local field decreases on entering the
superconducting state for A ($H\parallel\mba$), and the opposite
occurs for B ($H\parallel\mbbp$). Note that the relative change is
much smaller for B. Calibrations of the applied field were
determined to better than 10 parts per million (ppm) by
measurements of the $^{63}$Cu (in the coil) and $^3$He (in the
mixture in the vicinity of the coil) resonances. The
demagnetization field, arising from screening currents in the
superconducting state was determined to be less than 100ppm from
\C\ spectroscopy in sample B. The effect of temperature and field
is illustrated in Fig. \ref{fig:seShift}b. Note that no deviation
from the normal state shift is seen for $H=40$kOe.

\begin{figure}
\includegraphics[width=3in]{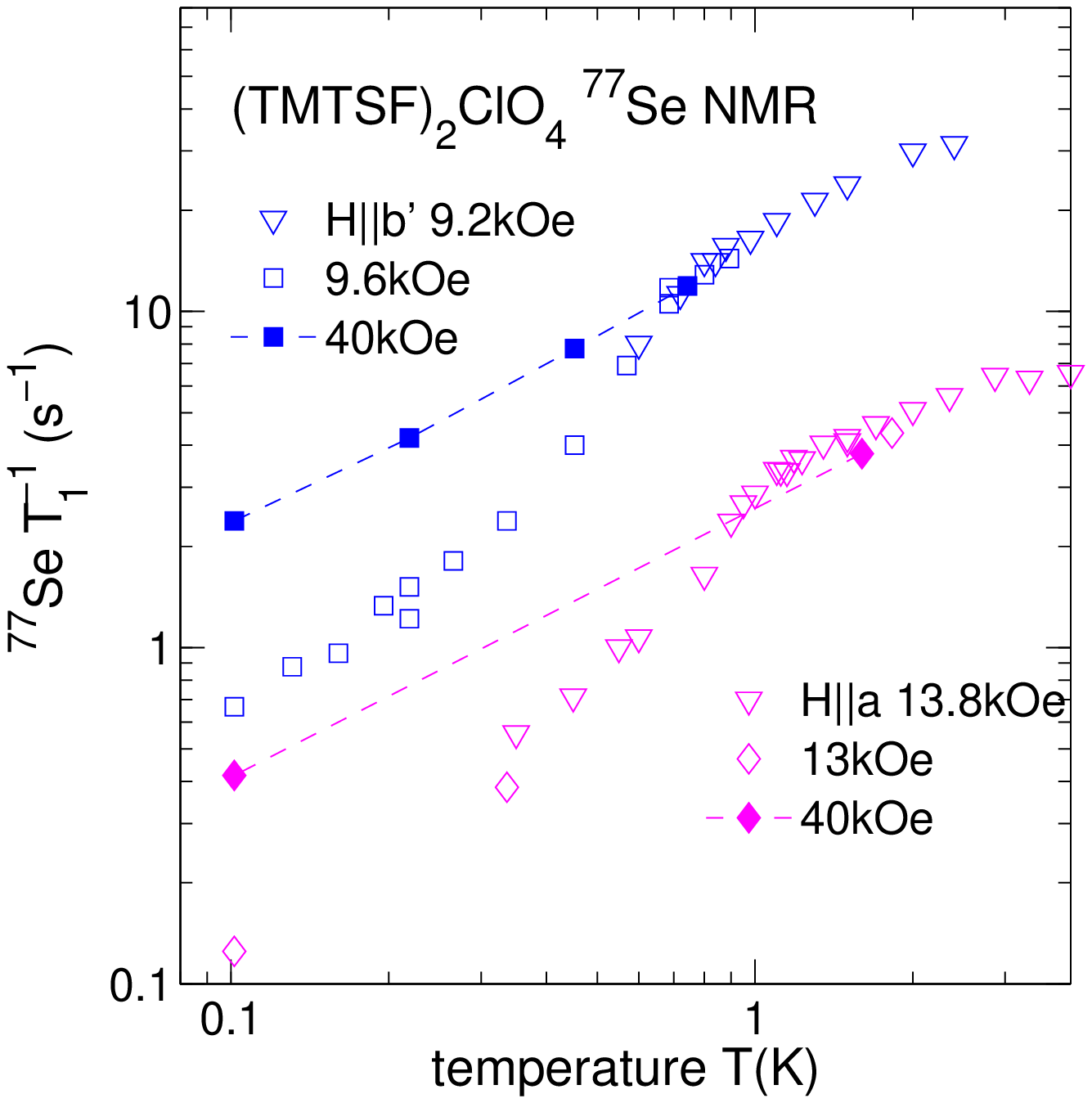}
\caption{\iTone vs. $T$ for (TMTSF)$_2$ClO$_4$, for
$\mathbf{H}\parallel\mbbp$ and $\mathbf{H}\parallel\mba$. Included
are results from prior measurements using a single-shot $^3$He
cryostat ($\triangledown$) \cite{Shinagawa:2006}, and results from
the current experiment.} \label{fig:iT1} \end{figure}

In Fig. \ref{fig:iT1}, we show the temperature dependence of
\iTone for both samples. The data collected at low field (open
symbols, see caption) exhibit a change of slope associated with
the superconductivity. No signature for superconductivity is
apparent for the data collected with $H=40$kOe (closed symbols),
which is close to the values for \hc2 reported elsewhere
\cite{Oh:2004}. Interpreting the change in slope as $T_c(H)$, we
obtain values for the critical field lower than reported in Ref.
\cite{Oh:2004} in both cases. We infer from the weak temperature
dependence of \iTone\ below $T\approx200$mK, that there is a
nonzero density of states at the Fermi level in at least part of
both samples at the lowest temperatures measured. If we were to
attribute the low temperature relaxation to a normal state
fraction phase segregated from the part that is superconducting,
then 30\% is the assigned fraction in the normal state.

The hyperfine coupling is nearly uniaxial: the dominant
contribution is a $p_z$ orbital originating at the Se sites
\cite{Takigawa:1986,Zhang:2005}. The normal state paramagnetic
shift is given by
\begin{subequations} \label{eqarray:hyperfinefields}
 \begin{eqnarray}
K&=&K_{iso}+K_{ax}(3\cos^2\theta-1),\\
K_{iso}&=&3.9(10)^{-4},\\
K_{ax}&=&10.5(10)^{-4}. %
\end{eqnarray}\end{subequations}
Thus, in low fields and for $T\ll T_c(H\to 0)$, we expect a change
for $\delta K_s\sim -700$ppm $(\theta=\pi/2,
\mathbf{H}\parallel\mbbp$), and $\delta K_s^a\sim +2500$ppm
($\theta=0, \mathbf{H}\parallel\mba$) for a superconductor with
singlet spin pairing. From Fig.~\ref{fig:seShift}, we observe
changes smaller than this using magnetic fields just less than
$10$KOe: $\delta K_s^b=-275$ ppm and $\delta K_s^a=+1500$ ppm,
respectively. In the first case, the observed value is a little
less than half what is expected for a singlet superconductor in
the small field, zero temperature limit. In the second case, it is
a little more. Unequivocally, $\chi_s$ is reduced in the
superconducting state. Further, the opposing signs of the change
are consistent with the known hyperfine couplings.

That $\chi_s$ does not completely vanish is not surprising when
compared to the measurements of \iTone. And attributing the
relaxation for $T\to0$ to hyperfine fields is confirmed by
comparing the rates at low temperatures in the \se\ and \C\
nuclei. Still, the character of the hyperfine fields is unknown.
For example, assuming that it arises from quasiparticles, it could
originate from the existence of a volume fraction in the normal
state, phase segregated from the superconducting portion. Another
possibility is a field- or disorder-induced density of states at
the Fermi energy. The observation of nearly single-exponential
relaxation at low temperatures for both samples speaks against
macroscopic phase-segregation.

\begin{figure}
\includegraphics[width=3in]{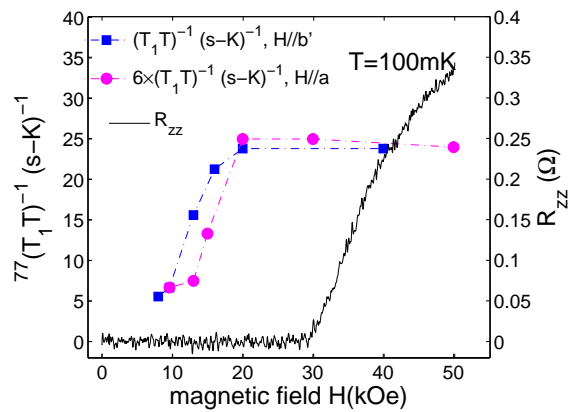}
\caption{\se \iTone vs. $H$ at $T=100$mK (left axis). Also shown
is the interlayer resistance
$R_{zz}(\mathbf{H}\parallel\mathbf{a})$ over a similar range of
fields (right-hand axis).} \label{fig:RzzH}
\end{figure}

To explore further this issue we measured the \se\ \iTone for
varying magnetic fields at $T=100$mK. This is shown in
Fig.\ref{fig:RzzH} for both field directions, along $\mba$ and
along $\mbbp$. Shown also is the interlayer resistance,
$R_{zz}(T=100$mK) vs. $H (\parallel\mba)$ for sample A. What is
notable in the relaxation rate is the fairly sharp increase
between $10$ and $20$kOe. We will refer to the ``crossover'' field
as $H_s$; it is much less than estimates of \hc2$\approx 50$kOe,
or more \cite{Oh:2004}. The result here is no exception: {\it in
situ} interlayer resistance measurements deviate from an
undetectable resistance only for $H>30$kOe, and clearly the
effects of superconductivity are evident to fields exceeding
$50$kOe. Unfortunately, a similar measurement for sample B was
unreliable because of a missing contact. 

\begin{figure}
\includegraphics[width=3in]{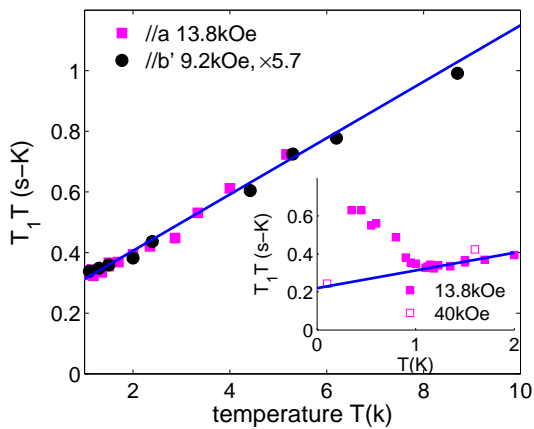}
\caption{\se\ $T_1T$ vs. $T$, demonstrating that the normal state
behavior follows closely to Eq. [\ref{eq:CW}]. Inset: An expanded
scale for $T<2$K, $H\parallel\mba$.} \label{fig:T1T}
\end{figure}

As superconductivity persists for $H>H_s$, we label the two
regimes as low-field SC ($H<H_s$, LSC) and high-field SC ($H>H_s$,
HSC). In the HSC regime, the relaxation rate \iTone\ is close to
the normal state value. The normal state behavior, shown in Fig.
\ref{fig:T1T} are remarkably well described by the empirical form
that is also characteristic of antiferromagnetic spin fluctuations
in 2D, \be T_1T=C(T+\Theta), \label{eq:CW} \ee with $C, \Theta$
constants \cite{Wu:2005:a}. Relaxation in the HSC regime is
similarly described, and should also result primarily from
hyperfine fields originating with quasiparticles.

The LSC regime exhibits a drop in $\chi_s$ that appears consistent
with a singlet superconductor. Impurity studies \cite{Joo:2005}
indicate a change in sign of the superconducting gap function over
the Fermi surface. And although the existence of nodes is
contradicted by thermal conductivity experiments
\cite{Belin:1997}, zero-field NMR relaxation in the
superconducting state \cite{Takigawa:1987} provides evidence for
the existence of nodes. Therefore, with the exception of the
results for thermal conductivity, the LSC is consistent with a
singlet state and nodes on the Fermi surface.

We are left to consider the nature of the HSC. We note that its
existence could account for the temperature independence of
$\chi_s$($H=14.3$kOe) reported in Ref.\cite{Lee:2001}. It is
unlikely to be filamentary for a number of reasons, most notably
that it is associated with a robust magnetic torque signal
\cite{Oh:2004}, and the zero resistance state is measured by many
laboratories without controversy. In that case, we have to take into
account the large \hc2. The suggestion it may be triplet followed
from this observation, and also because no phase transition to a
FFLO state was identified. This study calls that into question
because the apparent crossover at $H_s$ and $100$mK seems quite
sharp. Furthermore, there is evidence for a nonzero density of
states at the Fermi surface in the HSC regime, which is
qualitatively consistent with the FFLO. However, $H_{c2}(T\to0)$
exceeds estimates for the paramagnetical limit of the FFLO state
\cite{Lebed:1999}. An alternative to the FFLO state is a transition
to a triplet pairing state \cite{Shimahara:2000}; common to both
cases is the increase in the spin susceptibility of the
superconducting state, thus avoiding paramagnetic limiting.
Nevertheless, in considering these possibilities, it is not clear
why the spin lattice relaxation should be so close to the normal
state value as we observe. Consequently, a mapping of the phase
diagram , and more detailed NMR spectroscopy in the HSC regime
\cite{kakuyanagi:2005,Mitrovic:2006}, are necessary for a more
definitive description of the superconductivity for $H>H_s$.

In summary, it is established that the Bechgaard salt superconductor
\sclo\ is in the singlet state, most likely with gap nodes, at low
field. However, the $H-T$ phase diagram remains puzzling:
spin-lattice relaxation measurements give evidence for a sharp
crossover or phase transition at a field $H_s$ {\it within} the
superconducting state. We note that for the sample aligned
$H\parallel\mathbf b'$, assigning the steep increase in $T_1^{-1}$
near $H=H_s$ to hyperfine field fluctuations is verified by
comparing to the spin-lattice relaxation of $^{13}$C in the same
regime, whereas a similar check was not possible for the sample
aligned $H\parallel\mathbf a$. The nature of the HSC regime is
unknown and we consider the possibility that it is a transition to
an inhomogeneous FFLO state or a triplet-paired state. Confirmation
of the phase transition and the associated mapping of the phase
diagram, together with NMR spectroscopic information in the
high-field regime is necessary to clarify which of these
possibilities is the correct one, or whether the large spin
susceptibility in the HSC regime occurs for a different reason.


\small ACKNOWLEDGEMENTS. \normalsize The work was supported in part
by the NSF under grant no. DMR-0520552 (sb). We would like to
acknowledge discussions with P. M. Chaikin, A. Lebed, Y. Maeno, K.
Maki, and M. J. Naughton, and we thank M. J. Naughton for sharing
with us his results prior to publication.


\end{document}